\documentclass{PoS}
\usepackage{graphicx}
\usepackage{eurosym}

\title{Very-High-Energy gamma-ray astronomy with the ALTO observatory}

\ShortTitle{The ALTO gamma-ray observatory}

\author{ Yvonne Becherini$^{1}$, \speaker{Satyendra Thoudam}$^1$, Michael Punch$^{1,2}$, Jean-Pierre Ernenwein$^{3}$\\
        $^{1}$Department of Physics and Electrical Engineering, Linnaeus University, 35195 V{\"a}xj{\"o}, Sweden\\
        $^{2}$APC, Univ Paris Diderot, CNRS/IN2P3, CEA/lrfu, Obs de Paris, Sorbonne Paris Cit{\'e}, France\\
        $^{3}$Aix Marseille Univ, CNRS/IN2P3, CPPM, Marseille, France \\
        E-mail: \email{yvonne.becherini@lnu.se}}

\abstract{ALTO is a concept/project in the exploratory phase since 2013 aiming to build a wide-field VHE gamma-ray observatory at very high altitude in the Southern hemisphere. 
  The operation of such an observatory will complement the Northern hemisphere observations performed by HAWC and will make possible the exploration of the central region of our Galaxy and the hunt for PeVatrons,
  and to search for extended Galactic objects such as the Vela Supernova Remnant and the Fermi bubbles. 
 The ALTO project is aiming for a substantial improvement of the Water Cherenkov Detection Technique by increasing the altitude of the observatory in order to lower the energy threshold,  
 by using a layer of scintillator below the water tank to optimize the signal over background discrimination, by minimizing the size of the tanks and having a more compact array 
 to sample the air-shower footprints with better precision, and by using precise electronics which will provide time-stamped waveforms to improve the angular and energy resolution.  
 ALTO is designed to have as low an energy threshold as possible so as to act as a fast trigger alert to other observatories -- primarily to the Southern part of CTA -- for transient Galactic and extra-galactic phenomena.  
 The wide field-of-view resulting from the detection technique allows the survey of a large portion of the sky continuously, thus giving the possibility to access emission from Gamma-Ray Bursts, 
 Active Galactic Nuclei and X-ray binary flares, and extended emissions of both Galactic (Vela SNR, Fermi bubbles) and extra-galactic (AGN radio lobes) origin. The ALTO observatory will be composed of  
 about a thousand detection units, each of which consists of a Water Cherenkov Detector positioned above a liquid Scintillation Detector, distributed within an area of about $200\;\rm m$ in diameter. 
 The project is in the design study phase which is soon to be followed by a prototyping phase. The ALTO concept, design study and expected sensitivity together with the prototype status and plans 
 for final deployment in the Southern hemisphere will be the subjects of this presentation.}

\FullConference{35th International Cosmic Ray Conference --- ICRC2017\\
		10--20 July, 2017\\
		Bexco, Busan, Korea}

\begin{document}

\section{The aim of the ALTO project}
ALTO is a project to build a wide-field Very-High-Energy (VHE, $100\textrm{GeV}<E<100$TeV)
gamma-ray observatory at high-altitude in the Southern hemisphere,
see also \cite{ALTO} for updated information on the project. The
location in the Southern hemisphere is motivated primarily by the
privileged position for observing the central region of our Galaxy
and by the fact that another observatory of the same technological
``family'', the High Altitude Water Cherenkov project (HAWC) is
currently taking data in the Northern hemisphere near Puebla in Mexico,
see \cite{HAWC}. 

Being positioned in the Northern hemisphere, the HAWC observatory
can only marginally cover the Southern sky. With ALTO being installed
the Southern hemisphere, it will be possible to continuously observe
the centre of our Galaxy and, as an example, to monitor continuously
one of the most interesting and nearby Active Galactic Nuclei (AGN),
Centaurus A. Thus, with ALTO it will be possible to study the VHE
gamma-ray emission of astrophysical sources in our Galaxy and investigate
the origin of cosmic-rays from both Galactic and extra-galactic accelerators. 

The operation of ALTO in the Southern hemisphere will both complement
the VHE observations by HAWC, and will act as an all-sky monitor and
VHE alert system for the Southern part of the future Cherenkov Telescope
Array (CTA) observatory, see \cite{CTA} and \cite{CTAConcept}. 

ALTO will be constructed to study cosmic VHE gamma-rays above the
energy of a few hundred GeV, which bear witness to the most powerful
phenomena in the Universe since they originate in relativistic processes
in the vicinity of a wide variety of astrophysical sources such as
black holes, Pulsars, Supernova Remnants, Active Galactic Nuclei. 

\begin{figure}[t]
\begin{centering}
  \includegraphics[width=15cm]{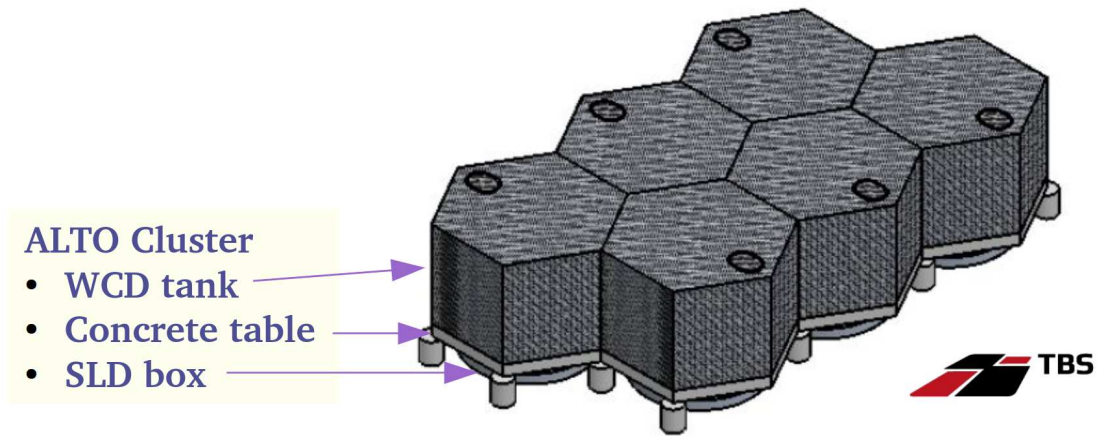}
  \caption{An ALTO cluster composed by six detection units.
    A unit is composed by a water Cherenkov detector (WCD) supported by a ``table'' of concrete.
    Under the concrete layer there is a ``muon detector'' 
    composed by a liquid scintillator tank (SLD; Scintillator Layer Detector).
    Detailed mechanical design of ALTO units is carried out by TBS Yard AB, see \cite{TBS}. }
  \label{ALTOCluster}  
\end{centering}
\end{figure}

The detection principle of ALTO is based on the sampling of the particle
content in the cascades generated by the interactions of gamma-rays
and cosmic-rays at the top of the atmosphere, where only the part of the cascade developed
just above the instrument is detected. An ALTO unit is composed by an hexagonal 
water Cherenkov tank, 2.5m-high and 3.6m-wide, supported
by a concrete structure; below the concrete layer a thin scintillator tank serves as
``muon detector''. The full ALTO array is planned to be composed of more than 1200 units
disposed in a circular area of $80\;\rm m$ of radius. The detection units are grouped
in ``clusters'' of 6 units whose signals are processed by a common electronics readout unit, see Fig. \ref{ALTOCluster},
and will also have in common solar panels and battery, plus the communication to central control by fibre.
In the current Monte Carlo simulations, each water tank is equipped with a single optical module 
containing a 10-inch photo-multiplier (PMT), while the scintillator tank is equipped with a 3-inch PMT.

In order to achieve as low an energy threshold as possible, the detectors
must be installed at very high altitude so as to reach the point at
which the particle development in the shower is closer to maximum. Candidate
sites for installation are in the regions of the Atacama desert in
Chile or near to San Antonio de Los Cobres in Argentina. Both sites
can offer an altitude of $5.1\;\rm km$ and are already equipped with excellent
scientific infrastructure, see \cite{ALMA}, \cite{QUBIC} and \cite{LLAMA},
so that ALTO will also benefit from this proximity. 

The advantages of the particle shower sampling technique for VHE gamma-ray
astronomy, besides being a quite inexpensive technology, are the possibility
to monitor a large portion of the sky simultaneously, the solid angle
being about $2\;\rm sr$\footnote{The exact value of the visible portion of the sky depends on the final
Zenith angle analysis cuts applied.}, and the possibility to observe the VHE Universe 24h per day with
no dead-time. The only physical limitation of the particle detection
technique is the constrained capability in direction reconstruction
accuracy with respect to IACT arrays, which sample the whole shower
geometry through the Imaging Atmospheric Cherenkov technique approach.
Another important parameter is, of course, the background rejection
capability, and this is something we wish to significantly improve
with ALTO.

When constructing such a research infrastructure in a such a remote
location which should take data for at least 30 years, several important
points have to be carefully considered. First of all the transport
and the installation of the infrastructure should be as simple as
possible so as to limit human stay at high altitude and costs. Secondly,
ALTO is carefully designed so as to have low maintenance and operation
costs, especially compared to IACT telescope arrays. Once installed,
ALTO has no moving parts, and there are no worries about the sunlight
destroying the equipment. Furthermore, there are no mirrors or entrance
windows which can be degraded by dust or UV light. Hence, for particle
shower detectors there is no need for a large crew doing maintenance
and repairs on the site, and this of course helps to minimize the
operation costs of the project. The current estimate of the total
cost of the full deployment of ALTO is of the order of $20\rm M$\euro, excluding
salaries.

The data generated by ALTO will follow the successful approach of the Fermi-LAT mission
of making the data available to the whole scientific community
possibly after a short proprietary period, see \cite{FERMI}, and so will be an open
observatory.

\section{Science with ALTO}

As mentioned above, ALTO belongs to the family of particle detectors
and is currently being optimised to provide a useful high-energy trigger
for CTA for transient events, from the ultra-short Gamma-ray Bursts
up to flaring X-ray binaries and Active Galactic Nuclei, to allow
studies of the VHE gamma-ray counterparts to the catastrophic and
powerful gravitational wave events seen by LIGO/Virgo (see \cite{LIGO}),
and, more generally, to be a key infrastructure in multi-messenger
astronomy.

The strength of the project lies in the possibility to monitor the
VHE southern gamma-ray sky continuously, giving the unique possibility
to have a multi-year view of the emission of southern objects with
no gaps in the lightcurve. Astrophysical sources will be visible once
per day during the time interval they transit through the ALTO Field-Of-View
(FoV) and with an optimized sensitivity plus favourable source spectra,
the transit in the FoV will be sufficient to detect a change in flux
and give an alert to the other observatories.

An infrastructure such as ALTO would catch Gamma-Ray Bursts (GRBs)
occurring in the large Field-Of-View (FoV) if their spectra are sufficiently
hard that they reach the ALTO energy range, would catch the flaring
episodes from AGN and X-ray binaries and would perform spectral measurements
up to their high-end. For GRBs and other transients such as gravitational
waves or cosmic neutrinos detected by other observatories, ALTO will
even be able to search in past data for counterparts, so greatly increasing
the multi-messenger potential.

In addition, it will be possible to search for extended emission from
Galactic objects, some of which are too large for IACT FoVs, such as
the Vela SNR and the Fermi bubbles, and in the extra-galactic domain,
to investigate the emission from AGN radio lobes. 

Nominally, VHE gamma-ray particle detectors can detect events up to
few hundreds of TeV, therefore it will be possible with ALTO to search
for new southern-sky PeVatrons, see \cite{HESS}. 

Given the high flux of charged cosmic rays detected, with dedicated
analyses it will be possible to determine and study the composition
in the range  $300\;\rm GeV$ -- $100\;\rm TeV$ and map large-scale structures in
the sky, especially the mysterious and intriguing cosmic-ray anisotropy
seen by MILAGRO \cite{MILAGRO}, Icecap \cite{ICECUBE} and Tibet-III
\cite{Tibet}. 

Other important scientific topics include the indirect detection of
Dark Matter, e.g\@. in Dwarf Galaxies; estimation of the Extra-galactic
Background Light from energy-dependent absorption of spectra of AGN;
search for Lorentz-Invariance violation from measurements of flares
of AGN; search for axion-like particles from distant AGN.
 
\section{Hybrid particle detection: Water Cherenkov and ``Muon-tagging'' Scintillators}

ALTO is a hybrid detector which aims to improve the current Water
Cherenkov technique used in HAWC on some key points: the altitude of the observatory,
the use of a layer of scintillator below the water tank, the construction
of smaller tanks, and the use of more precise electronics and time-stamping.
In addition, we will describe here an idea which could help in keeping
the costs of the observatory low. 

\textbf{Altitude of the observatory.} By increasing the altitude of
the observatory location with respect to HAWC one
can approach the point at which the number of particles in the gamma-ray
induced shower is maximum, and this gives the possibility to decrease
the energy threshold of the gamma-ray events. In particular we plan
to increase the altitude up to $5.1\;\rm km$ which allows a decrease of 40\%
in the energy threshold with respect to HAWC. 

\textbf{Background rejection.} In our current ALTO design, we have
added of layer of scintillator below the water tank. This is very
important in order to be able to ``tag'' the passage of muons, which
are the almost unambiguous signature of the nature of the particle
cascade, as background proton-initiated showers are muon-rich. The
implementation of this new ``muon-tagging'' allows to reach an increased
signal over background discrimination by analysis, and thus allows
an increase of the sensitivity of the detection technique. One important
consideration is the type of liquid scintillator to be used, given
that ALTO will be in a remote location, and that the flora and the
fauna should be protected. One liquid scintillator which is under
consideration is a non-toxic oil with some non-toxic dopants which
is relatively cheap, while being effective and environmentally friendly. 

\textbf{Tank design. }The construction of smaller tanks with respect
to HAWC will allow to have a finer-grained view of the shower particles
on the ground which helps in the reconstruction of the arrival direction
of the incoming event and in the background rejection. The current
design of the ALTO water tank is hexagonal, so as to be close-packed. 

\textbf{Improvements in front-end electronics and timing.} ASIC Analogue
Memories allow to store an electronic signal for a short time in a
circular buffer (Switched-Capacitor Array, SCA), only reading it out
if a trigger condition has passed. 
For systems where the read-out is not continuous, but happens in ``windows''
at a low rate, this allows the digitization to be done at a slower
rate, avoiding the need for expensive, fast, and power-hungry flash
Analogue to Digital Converters (ADCs). 
For ALTO we propose to take advantage of the improvements in integrated
electronics which have been made in the last 10--15 years, in particular
for the aspects of the Analogue memories and the timing distribution.
One possible solution for the full ALTO array is to use the NECTAr
ASIC used in CTA, see \cite{Delagnes} and \cite{Glicen}.

\textbf{ANTARES optical modules.} The Antares underwater neutrino
telescope is reaching the end of its operation, see \cite{ANTARES}.
The 12 lines of the detector have been taking data from 2008 and will
be recovered soon. The Antares Collaboration is accepting proposals
on how to recycle its 900 optical modules, each of which contains
a 10-inch Hamamatsu photo-multiplier.
One possibility is that these are reused in ALTO.
 
\begin{figure}[t]
\begin{centering}
  \includegraphics[width=12cm]{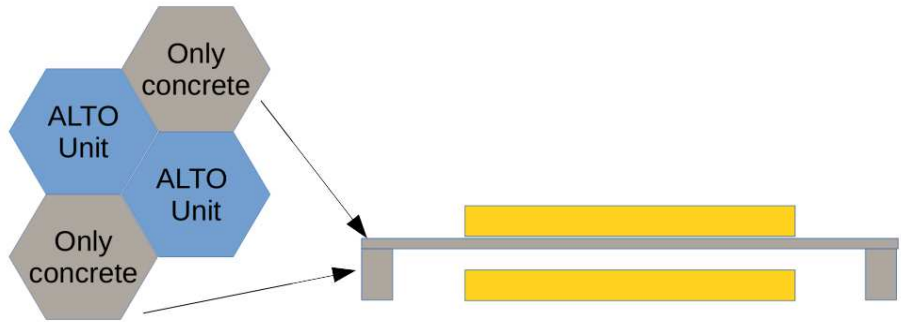}
  \caption{Configuration of the ALTO prototyping facility at Linnaeus University in Sweden.
    The concrete ``table'' will be built for 4 units, while only two ALTO units will be installed for the moment.
    The remaining two empty slots will be equipped with several additional scintillator boxes
    recycled from an on-board air shower detector array used by the ANTARES experiment for calibration purposes, shown in yellow.
   }
    \label{ALTOProto}  
\end{centering}
\end{figure}

\section{The ALTO prototype at Linnaeus University}

After almost two years of Monte Carlo simulations and after several design iterations, we have converged to a design that we have chosen for the prototyping phase, see \cite{Satyendra}.
In 2017, a prototype of two ALTO units is being installed at the Linnaeus University campus in V{\"a}xj{\"o}, Sweden.
The detailed design of the units is currently being finalized in collaboration with TBS Yard AB, see \cite{TBS}, the Swedish company in charge of the construction. 
The full prototype setup will consist of two full ALTO units and several smaller scintillator boxes, see Fig. \ref{ALTOProto}. 
These additional scintillator boxes are recycled from an on-board air shower detector array used by the ANTARES experiment for calibration purposes, see \cite{ANTARES}, and will be used in different modes
firstly in order to study the detector responses to muons with known tracks, and secondly for timing calibration tests. 
During the prototyping phase, several different PMT solutions will be tested in parallel, and we will use an electronics similar to NECTAr, called WaveCatcher, see \cite{WaveCatcher}, for the read-out.
From the operation of the prototype we expect to measure the rates in the Water Cherenkov and the scintillator channels, so that this measured rate can be compared with simulations and extrapolated to the rate at $5.1\;\rm km$ altitude,
and we will also gain experience in installation and operation.

\section{Next steps}
Depending on the outcome of the prototype phase, we will see if the design will need adjustments. 
The next step of the ALTO project, possibly in 2019, will be to install a full cluster of 6 detection units directly on the chosen site in the Southern hemisphere.
This will allow to do further testing on the capability and the ease of building such detectors at high altitude. 

\section{Conclusions}

A wide-field VHE gamma-ray observatory is missing in the Southern hemisphere.
The presence of such an observatory continuously looking at the Southern sky with no dead-time is a tremendous opportunity for gaining access to the highest-energy phenomena at work in the Universe.
ALTO can operate as a stand-alone observatory as well as a high-energy trigger alert system for the Cherenkov Telescope Array (CTA) and as a key infrastructure for multi-messenger astronomy.
Both detailed Monte Carlo simulations and initial prototype designs are well-advanced, with concrete comparisons of the two expected soon, for a solid extrapolation to the full future ALTO observatory.

\section{Acknowledgements}

The ALTO project is being supported by the following Swedish private foundations or public institutes: the Crafoord Foundation, the Foundation in memory of Lars Hierta,
the Magnus Bergvall's Foundation, the Crafoord stipendium of the Royal Swedish Academy of Sciences (KVA),
the M{\"a}rta and Erik Holmberg Endowment of the Royal Physiographic Society in Lund, the L{\"a}ngmanska kulturfonden, the Helge Ax:son Johnson's Foundation
and Linnaeus University. We also thank the Swedish National Infrastructure for Computing (SNIC) at Lunarc (Lund, Sweden).\\
We would like to thank Bertrand Vallage from CEA/Saclay (France) for providing us with two ANTARES optical modules.
Thanks also to Staffan Carius, Dean of the Faculty of Technology at Linnaeus University, for all the local support for the project.


\begin{thebibliography}{99}

\bibitem{ALTO}The ALTO website: http://alto-gamma-ray-observatory.org
\bibitem{HAWC}The HAWC website: http://www.hawc-observatory.org/
\bibitem{CTA} The CTA website: http://cta-observatory.org/
\bibitem{CTAConcept} Acharya, B. S. et al., Introducing the CTA concept, Astropart. Phys., 43, 3-18, (2013)
\bibitem{ALMA} The ALMA website: http://www.almaobservatory.org/
\bibitem{QUBIC} The QUBIC website: http://qubic.in2p3.fr/QUBIC/Home.html
\bibitem{LLAMA} The LLAMA website: http://www.iar.unlp.edu.ar/llama-web/english.html
\bibitem{FERMI}  The Fermi mission website: https://fermi.gsfc.nasa.gov/
\bibitem{LIGO} LIGO Scientific Collaboration website: http://ligo.org
\bibitem{HESS} H.E.S.S. coll., Nature 531, 476-479 (2016) 
\bibitem{MILAGRO} MILAGRO papers, http://hawc.pa.msu.edu/milagro$\_$papers/papers.html
\bibitem{ICECUBE} The IceCube neutrino observatory website , https://icecube.wisc.edu/
\bibitem{Tibet} Tibet-III, http://www.icrr.u-tokyo.ac.jp/em/
\bibitem{Delagnes} E. Delagnes, J. Bolmont et al. Proc. IEEE NSS/MIC Conference, Valencia, Spain, 1457 (2011)
\bibitem{Glicen} J.F. Glicenstein et al., the CTA Consortium, Proc. 33rd ICRC, Rio de Janeiro, (2013)
\bibitem{ANTARES} The ANTARES website: http://antares.in2p3.fr/
\bibitem{Satyendra} S. Thoudam, Y. Becherini, M. Punch, Simulations study for the proposed wide field-of-view gamma-ray detector array ALTO, at this conference
\bibitem{TBS} http://www.tbsab.com/
\bibitem{WaveCatcher} D. Breton, E. Delagnes, J. Maalmi et al., The WaveCatcher family of SCA-based 12-bit 3.2-GS/s fast digitizers, DOI: 10.1109/RTC.2014.7097545

\end{thebibliography}
\end{document}